\documentclass[english,letterpaper,twocolumn,showpacs,pra,aps]{revtex4}
\usepackage{graphicx}
\usepackage{amssymb}

\makeatletter

\large  

\input epsf

\makeatother

\usepackage{babel}
\makeatother
\begin{document}

\title{Casimir effect in a one-dimensional gas of free fermions}

\author{Eugene B. Kolomeisky,$^{1}$ Joseph P. Straley,$^{2}$ and Michael
Timmins$^{1}$}

\affiliation{$^{1}$Department of Physics, University of Virginia, P. O. Box 400714,
Charlottesville, Virginia 22904-4714, USA\\
$^{{2}}$Department of Physics and Astronomy, University of Kentucky,
Lexington, Kentucky 40506-0055, USA}

\begin{abstract}
We compute an analog Casimir effect in a one-dimensional spinless Luttinger liquid confined to a segment  in the presence 
of a nearly-impenetrable partition dividing the segment into two 
compartments.  The
Casimir interaction is found to be a bounded piecewise-continuous 
oscillatory function whose maxima are points of force discontinuity and 
correspond to resonant tunneling across the partition. 
The well-known regularization-based results are reproduced by the lower envelope of this function, which  corresponds to an approximation that ignores the rather large oscillations due to particle discreteness.  These macroscopic conclusions are tested and confirmed via a rigorous analysis of the Casimir effect in an exactly-solvable model of a one-dimensional non-relativistic spinless gas of  
free fermions, thus resolving an objection that has been raised by Volovik (2003).  
 Additionally we confirm the result of a recent calculation 
which employed an effective
low-energy theory with a cutoff to find the Casimir interaction 
between two strong well-separated impurities placed in a Luttinger liquid.
\end{abstract}


\maketitle

\section{Introduction}

Casimir interactions are macroscopic manifestations of the vacuum
zero-point energy brought about by introduction of boundaries or changes in the topology
of space.  These modify the spectrum of zero-point
fluctuations from their form for free space, which gives rise to
experimentally detectable forces \cite{experiment}.
Although originally derived as an attractive interaction 
between perfectly conductive parallel plates induced by vacuum 
fluctuations of the electromagnetic field \cite{Casimir}, similar 
effects are present for other fields, boundary conditions, 
and geometries \cite{CasReviews}. Analogous phenomena take 
place in classical physics due to equilibrium thermal 
fluctuations \cite{Kardar}.

Although important in their own right, the existence of Casimir interactions has a series of 
implications spanning various areas of science \cite{PT}. For instance, the van der Waals 
counterparts of these forces are crucial in understanding the phenomena of 
wetting \cite{deGennes}. In nanoscale devices Casimir attractions may lead to an undesirable 
effect of "stiction" of closely positioned parts of an apparatus \cite{stick}. Casimir forces 
can also be made useful; the first micromachines employing them were recently built \cite{micro}. Experimental measurements of the Casimir force provide constraints on the existence of extra space dimensions and fundamental physics beyond the standard model \cite{extra}, while dynamic versions of the phenomenon \cite{Kardar} are related to the Hawking radiation from black holes and Unruh effect of radiation from accelerating objects. 

The Casimir interaction $E_{C}$ can be defined as a difference between the vacuum 
energy $E_{vac}$ of the system constrained by the boundaries or changes in topology and 
the vacuum energy
of free space:
\begin{equation}
\label{Casimirdefinition}
E_{C} = E_{vac~constrained} - E_{vac~free}
\end{equation}
The vacuum energies, in turn, are calculated from a harmonic field 
theory (such as quantum electrodynamics in the case of the electromagnetic Casimir effect), 
thus implying that $E_{vac}$ is the sum of zero-point energies of a collection of simple harmonic 
oscillators with a spectrum $\omega(\textbf{k}) = c |\textbf{k}|$:
\begin{equation}
\label{vacenergy}
E_{vac} = \frac{1}{2} \sum \hbar \omega(\mathbf{k})
\end{equation}
where the summation is performed over all branches of the spectrum and over all allowed wave vectors $\textbf{k}$. The Casimir interaction arises due to a difference between the allowed set of wave vectors $\textbf{k}$ of the constrained system and that of free space.

In any realization of the Casimir effect, the boundaries are made of real materials that are transparent at extremely
short wavelengths, so that the high-energy parts of Eq. (\ref{vacenergy}) are insensitive to the  presence of the boundaries thus implying a cancellation in Eq.(\ref{Casimirdefinition}).   For practical calculations, one studies the case of an  impenetrable partition and applies a regularization scheme
which amounts to the introduction of a function  
$f(\lambda, \textbf{k})$ such that $f(\lambda \rightarrow 0, \textbf{k}) = \omega(\textbf{k})$. 
The role of the parameter $\lambda$ consists in suppressing the large-$\textbf{k}$ contributions 
into the vacuum energies to guarantee their convergence. Then the Casimir interaction is found 
from Eqs. (\ref{Casimirdefinition}) and (\ref{vacenergy}) by taking the 
limit $\lambda \rightarrow 0$. 

For suitable choices of the function $f(\lambda,  \textbf{k})$, the outcome of this procedure
gives a finite result that depends only on Planck's constant, the speed of light and macroscopic length scale(s) indicating
that we are dealing with a long-wavelength phenomenon; the same result has been obtained with 
different regularization schemes, so that it has a universal character.  
There is general agreement on the correctness of the result, and yet the route to it is subtle;
it is certainly possible to get other (physically meaningless) evaluations of the sum 
(\ref{vacenergy}).
It is a little surprising that in the cancellations that provide a finite result for
Eq. (\ref{Casimirdefinition}), no trace of the possible differences between long-wavelength
and short-wavelength physics survives, even though  this would significantly affect the sum (\ref{vacenergy}).

To shed some light on this and other issues having their roots in diverging vacuum energy, Volovik \cite{Volovik1} advocated an approach consisting in studying analog effects in condensed matter systems. 
Here the role of the physical vacuum is played by the ground state of a condensed matter system, and Casimir-type effects occur whenever the ground state is disturbed by boundaries or changes of space topology. 
The ground-state energy density of a condensed matter system is finite, and neither the harmonic approximation nor regularization are needed to calculate the Casimir effect. This is because the underlying microscopic physics is governed by the well-established quantum mechanics of non-relativistic particles interacting via non-singular interactions. 
As a result any quantity of interest can be calculated, at least in principle.   
When the low-energy physics of a candidate condensed matter system is described by a harmonic 
field theory, we ought to find the same interaction as is predicted
by a regularization scheme.

There are numerous condensed matter systems whose low-energy physics is described by harmonic 
field theories. Superfluid $^{4}He$ in its ground state is the simplest analog to the vacuum 
of quantum electrodynamics, where the role of photons is played by phonons propagating with 
sound velocity $c$. However the dispersion law $\omega = c |\textbf{k}|$ is only applicable 
as long as the phonon wavelength $2\pi/|\textbf{k}|$ is significantly larger than interatomic 
distance. 

\begin{figure}
\includegraphics[width=1.2\columnwidth,keepaspectratio]{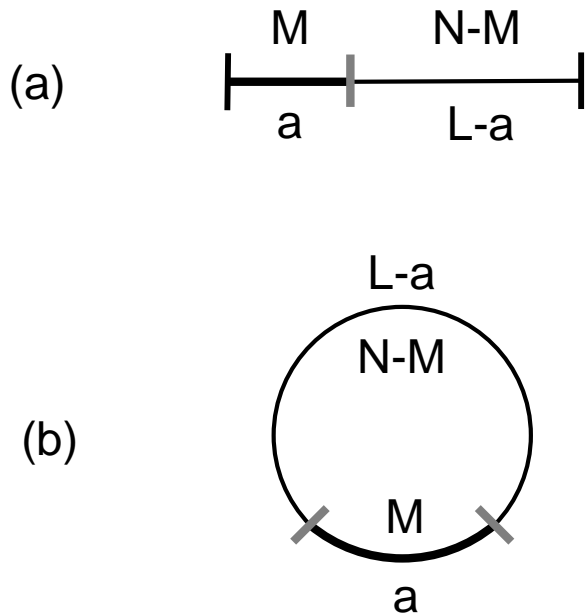} 
\caption{Geometries employed to study Casimir effect in one dimension: (a) box of length $L$ with $N$ interacting fermions or repulsive bosons and a nearly-impenetrable partition (shaded gray) placed a distance $a$ away from one of the box walls. For impenetrable box walls the ground-state energy of this system is identical to that shown in (b): $N$ particles in a ring of circumference $L$ with two nearly-impenetrable partitions. Regular and bold typefaces show correspondence between geometries (a) and (b).}
\end{figure}

Volovik provided an example of an exactly-solvable toy condensed matter 
system \cite{Volovik2} where the Casimir interaction seemed to disagree with the 
regularization-based conclusion (which is a one-dimensional counterpart of Casimir's original force between perfectly-conducting plates). He considered a one-dimensional gas of non-relativistic free 
spinless fermions in the geometry of Fig.1a, where $N$ particles are 
placed inside an impenetrable box of length $L$ with a nearly-impenetrable partition 
(shown in gray shade). Although this is a non-interacting system, nontrivial correlations 
are built in via the Pauli principle. The partition breaks the system into two 
nearly-impenetrable boxes of length $a$ and $L - a$; an infinitesimally small tunnelling 
transparency allows particle exchange between the boxes to minimize the energy. The number of 
fermions $M$ captured between one of the walls of the box and the partition can be found by 
minimizing the ground-state energy $E(M, a, N - M, L - a)$ with respect to $M$ under the 
constraint that $M$ can only take on integer values. As a result the $M(a)$ dependence has the 
form of a staircase with plateaus located at integer values of $M$. The inter-plateau 
transitions are connected via resonant tunneling since the ground state here is degenerate: 
$E(M, a, N - M, L - a) = E(M \pm 1, a, N - M \pm 1, L - a)$. Volovik explicitly demonstrated 
that at the points of degeneracy the force 
exerted on the partition, $-\partial E_{C}/\partial a$, undergoes a discontinuity whose 
magnitude is parametrically larger than that of the ordinary Casimir force implied by the
regularization-based result -- even in the limit of macroscopic occupancy of both boxes. 
He called this phenomenon the "mesoscopic" Casimir effect 
because "high-energy" physics (arising from the discreteness of the particles constituting the 
"vacuum") explicitly enters into the result even when the sizes of the boxes $a$ and $L - a$ 
are much larger than average interparticle spacing. A similar conclusion was reached for a 
one-dimensional model of "ultra-relativistic" spinless free fermions with a linear 
dispersion law \cite{Volovik3}.

More recently the Casimir interaction between two well-separated 
impurities immersed in a one-dimensional spinless quantum liquids was 
studied within an effective low-energy theory with a cutoff \cite{Zwerger} 
and in a lattice model \cite{lattice}.  A closely related problem is that of the interaction between two vacancies in the one-dimensional spin-1/2 antiferromagnetic Heisenberg model \cite{Anfuso}.   These studies are related to the $L \gg a$ limit of the ring geometry shown in Fig.1b in the 
presence of two barriers. It is easy to realize that when the 
barriers become barely penetrable the ground-state energy of any 
quantum liquid in the geometry of Fig.1a approaches that in Fig. 1b. 
Thus for non-interacting fermions and strong impurities it should be 
possible to compare the conclusions of Refs. \cite{Zwerger} and 
\cite{lattice} with Volovik's findings \cite{Volovik2}. 

It was found \cite{Zwerger} - \cite{Anfuso} that the Casimir 
interaction is a variable-sign oscillatory function of the impurity/vacancy
separation $a$, with maxima for the separations which permit resonant 
tunneling.
 The maxima of the Casimir energy represent the points 
of discontinuity of the slope. Qualitatively this agrees with 
Volovik's results \cite{Volovik2}.  We also note that an oscillatory interaction between various kinds of inhomogeneities immersed in an ideal three-dimensional Fermi gas was found by Bulgac \textit{et al.} \cite{Bulgac}.  

The goal of this paper is two-fold:  first contributions of  the effects of the particle discreteness into Casimir-type phenomena are computed  for an arbitrary spinless quantum liquid by employing the conventional macroscopic field-theoretical viewpoint.   Second, these phenomenological conclusions are verified through a rigorous microscopic analysis of the Casimir effect for spinless free fermions. As a result we will be able to: (i) establish a relationship between the ordinary and mesoscopic Casimir effects, and (ii) to link the macroscopic and microscopic approaches.
 
In addition to these issues of principle, the problem considered here 
may be of practical interest as the Casimir forces between impurities 
immersed in fermionic quantum wires realized with ultracold gases 
\cite{wire} might be experimentally measured in the future.   

Before proceeding we hasten to mention that the effects calculated in this work should not be confused with the 
relativistic fermionic Casimir effect (with bag-model boundary conditions on the surface),
which for parallel plates was first worked out by Johnson \cite{Johnson}.

The organization of this paper is as follows. In Section II we summarize 
conventional  predictions of the harmonic theory and explain 
how the complications coming from particle discreteness can be accounted for
phenomenologically within the harmonic theory.  
Section III turns to the non-interacting fermion system, showing 
how these results can be verified and confirmed via rigorous microscopic analysis.
Section IIIA deals with the case of a segment
whose ends are subject to either an arbitrary combination of the 
Dirichlet and Neumann or to periodic boundary 
conditions imposed on the many-body wavefunction. Here the Casimir effect manifests itself as a finite-size correction to the ground-state energy. The irrelevance of the effects 
of the particle discreteness in the segment or ring geometries allows 
us to probe only the continuum part of the Casimir force.  

A more complicated geometry where the effects of the particle 
discreteness come into play (a segment with nearly-impenetrable 
partition or two nearly-impenetrable partitions inserted into a ring) 
is studied in Sections IIIB and IIIC. The former deals with the limit when 
the particle discreteness continues to be ignored. In close 
correspondence with the conclusions of Section IIIA, we reproduce the 
results of the effective theory for the case of the Dirichlet boundary 
conditions enforced at the segment ends (which is equivalent to the 
case of two partitions inside a ring).  The effects of 
particle discreteness are incorporated in Section IIIC where macroscopic results of Section IIA are confirmed.   We conclude (Section IV) by discussing the correspondence between microscopic and macroscopic viewpoints and posing several questions for future study. 

\section{Macroscopic approach}

Volovik's fermion model exhibits large oscillations in the Casimir energy, which
arise from effects of particle discreteness.  In the following subsections
we show that these can be understood phenomenologically without resorting to microscopic analysis.
\subsection{Continuum results}

 We consider a generic
 harmonic spinless quantum  liquid in the geometry of Fig.1.
We begin with restating the well-known result \cite{Luttinger} that the 
long-wavelength low-energy dynamics of a translationally-invariant 
one-dimensional spinless quantum liquid comprised of interacting 
fermions or repulsive bosons of mass $m$ is governed by the harmonic 
(Luttinger liquid) theory with the Lagrangian
\begin{equation}
\label{action}
\mathcal{L} = {\frac{mn}{2}}\int dx\left (({\frac{\partial u}{\partial t}})^{2} -
c^{2}({\frac{\partial u}{\partial x}})^{2}\right )
\end{equation} 
where $x$ is the position, $t$ is the time, and $n$ is 
the number density. The field $u(x, t)$ represents displacement of 
the particles with respect to the case of equidistant positions. 
The first term of (\ref{action}) describes the kinetic energy while 
the second accumulates pertinent information about the interparticle 
interactions (and statistics, in case of fermions) parameterized by 
the sound velocity $c$. This latter quantity is determined by the 
ground-state 
energy; for spinless free fermions the sound velocity 
$c = \pi \hbar n/m$ is identical to the Fermi velocity. The 
description (\ref{action}) is applicable for spatial scales exceeding the
interparticle spacing $n^{-1}$, provided that the underlying 
interactions are sufficiently short-ranged and the 
strain is small ($|\partial u/\partial x| \ll 1$). 

The interactions due to finite size described above will have the  universal Casimir form but the boundary conditions also matter.  In cases of interest the
collective variable $u(x, t)$ satisfies either the Dirichlet 
condition $u = 0$, 
or the Neumann condition $\partial u/\partial x = 0$.
The former mimics the perfectly-conducting plate of the original Casimir 
calculation \cite{Casimir} while the latter corresponds to a 
plate of infinite magnetic permeability \cite{Boyer}. If in the geometry of Fig. 1a the
partition and the walls enforce "like" boundary conditions 
(i. e. either Dirichlet or Neumann at all boundaries), then 
for any smooth regularization scheme (and up to a constant) the 
Casimir interaction was found to be of the form \cite{Boyer}
\begin{equation}
\label{ocasimir}
E_{C, like} = - \frac{\pi \hbar c}{24} \left (\frac{1}{a} + \frac{1}{L - a}\right )
\end{equation} 
The case of mixed boundary conditions is particularly interesting because 
Boyer demonstrated \cite{Boyer} that for the geometry of Fig. 1a the Casimir 
interaction is repulsive
\begin{equation}
\label{unlike}
E_{C, unlike} = \frac{\pi \hbar c}{48} \left (\frac{1}{a} + \frac{1}{L - a}\right )
\end{equation} 
when the partition and the walls enforce "unlike" boundary conditions \cite{Boyer} (i.e. either Neumann at 
the walls and Dirichlet at the partition or Dirichlet at the walls and Neumann at the 
partition). 

For $L \rightarrow \infty$ Eq.(\ref{ocasimir}) gives $E_{C, like} = - \pi \hbar c/(24a)$ which 
is a well-known result \cite{Johnson}, \cite{NPB}, \cite{conformal}, while Eq.(\ref{unlike}) predicts $E_{C, 
unlike} = \pi \hbar c/(48a)$. Regardless of the boundary conditions employed, the 
$\hbar c /a$ behavior of the Casimir interaction $E_{C}$ can be understood via dimensional analysis. Indeed, since the effect is both quantum-mechanical and "relativistic", $E_{C}$ must depend on $a$, $\hbar$ and $c$. If we additionally assume that the effect is universal, then $E_{C}$ cannot depend on microscopic quantities such as the mass of the underlying particles or the wavevector cutoff $n$. Given $E_{C}$, $\hbar$, $c $ and $a$ one can form one and only one dimensionless combination $E_{C}a/\hbar c$ which implies $E_{C} = const (\hbar c /a)$ where the numerical constant is only determined by the boundary conditions. 

\subsection{Effects of particle discreteness}
 
The Lagrangian (\ref{action}) describes the elastic theory of a  one-dimensional solid.  This viewpoint is acceptable because despite the fact that the long-range translational order is destroyed by the zero-point motion \cite{Mermin}, algebraic correlations survive and typical strains remain small.   Quantum fluctuations are additionally suppressed under confinement as in geometries of Fig.1.  

A complication arises
in the discussion of the elastic theory of a solid, in which the field variable 
describes the displacements of discrete particles from equilibrium positions.  When
we change the length of a box containing a certain number of particles, we 
change both
the wavelengths of the excitations of the system and the equilibrium positions
themselves:  the solid experiences a strain, which changes the ground-state energy.  When the strain becomes large enough, it can be relaxed (as in geometries of Fig.1) by adding
or removing a particle, but this is a discrete effect.  A second complication is that 
the introduction of the boundaries will cause particle displacements, no matter 
where they are put, because the particles might be attracted or repelled by the boundaries.  
These effects can be described within the harmonic theory (\ref{action}), as we will now demonstrate.

The ground-state energy  of a quantum liquid in the geometry of Fig. 1a can be obtained by minimizing the sum of the ground-state energies of separate arbitrarily populated boxes  $E(M,a) + E(N-M,L-a)$ with respect to $M$ provided the latter is restricted to the integers.  In the macroscopic limit $M, N - M  \gg 1$ when the discreteness of the underlying particles can be at first neglected, this condition reduces to the statement of the equality of the chemical potentials on the two sides of the partition.  This in turn implies that  the bulk densities (and thus the sound velocities) on the two sides of the partition are the same.  The common density $n$ defines the reference zero-strain state, and the field variable $u$ in Eq.(\ref{action}) describes displacements away from corresponding interparticle spacing $n^{-1}$.  Although the chemical potentials on the two sides of the partition are equal, the forces are not since the partition is not in mechanical equilibrium.  The corresponding osmotic force on the partition is the continuum Casimir force implied  by Eq.(\ref{ocasimir}) and (\ref{unlike}). 

When the discreteness of the particles is taken into consideration, a particle exchange across the partition can only occur at  the points of degeneracy of the ground state.  As explained above generally this causes a strain which can be computed as follows.  The number of particles captured between $x = 0$ and $x = a$, namely $n(a - u(a,t) + u(0,t)) - \varphi$,  is constrained to be an integer $M$.  This leads to the boundary condition
\begin{equation}
\label{boundarycondition}
u(a,t) - u(0,t) = a-\frac{M+\varphi}{n}
\end{equation}
Here $\varphi$ is a correction of order unity due to the disturbance of the density profile away from uniformity caused by the ends of the interval.  This correction cannot be computed within the harmonic approach since the latter fails in the vicinity of the left wall  and the partition.

The boundary condition (\ref{boundarycondition}) implies that in general there will be a uniform static strain of the form
\begin{equation}
\label{strain}
du(x)/dx = \frac{u(a,t) - u(0,t)}{a} = 1 - \frac {M+\varphi}{na} 
\end{equation}
The system can also have time-dependent
motions in addition to the static strain; these are described by the 
wave equation  
$\partial^{2}u/\partial t^{2} - c^{2}\partial^{2}u/\partial x^{2}=0$.  

For example, if the field $u(x,t)$ obeys the Dirichlet boundary condition at the $x = 0$ wall, $u = 0$, then for $0\leqslant x \leqslant a$ the corresponding general solution to the wave equation satisfying the boundary condition (\ref{boundarycondition}) is given by
\begin{equation}
\label{configuration}
u(x,t) = (1-\frac{M+\varphi}{na})x + \Re \sum_{l=1}^{\infty}b_{l} e^{i \pi lct/a} \sin \frac{\pi lx}{a}
\end{equation}
where the $b_{l}$ are arbitrary complex coefficients.
The sum in Eq.(\ref{configuration}) represents the field configurations 
responsible for the continuum Casimir effect in an interval of length $a$ 
with Dirichlet-Dirichlet (DD) boundary conditions imposed at the segment ends.  Substituting the 
solution (\ref{configuration}) into the Lagrangian (\ref{action}) we find
\begin{equation}
\label{lagrangian}
\mathcal{L} = \mathcal{L}^{(DD)} - \frac{mc^{2}}{2na}(na - M - \varphi)^{2}
\end{equation}
where $\mathcal{L}^{(DD)}$ is the Lagrangian corresponding to the sum in Eq.(\ref{configuration}).  It is important to notice that the cross terms  occurring in the evaluation of $(\partial u/\partial x)^{2}$ vanish upon integration over $x$.  This implies that the term linear in $x$ and the sum in (\ref{configuration}) contribute independently to the ground-state energy.  Therefore the portion of the Casimir energy coming from the $0 \leqslant x \leqslant a$ modes is given by
\begin{equation}
\label{portion}
E^{(DD)} + \frac{mc^{2}}{2na}(na - M - \varphi)^{2} = - \frac{\pi \hbar c}{24a} + \frac{\pi \hbar c}{2ga}(na - M - \varphi)^{2}
\end{equation}  
where $E^{(DD)} = - \pi \hbar c/(24a)$ is the first term of Eq.(\ref{ocasimir}).   
The dimensionless parameter 
\begin{equation}
\label{exponent}
g = \frac{\pi \hbar n}{mc}
\end{equation}
in the second term of (\ref{portion}) can be recognized as the Luttinger liquid exponent governing the low-energy behavior of the correlation functions in one-dimensional harmonic liquids \cite{Luttinger};  the free-fermion case corresponds to $g = 1$.

For the geometry of Fig.1a there are two segments.  The portion of the Casimir energy 
coming from the modes in the segment $a < x \leqslant L$ is computed 
similarly.  Combining it with Eq.(\ref{portion})  we arrive at the expression
\begin{eqnarray}
\label{EDDDgeneral}
E_{C}^{(DDD)}& = &- \frac{\pi \hbar c}{24}\left ( \frac{1}{a}+\frac{1}{L-a}\right )\nonumber\\
&+& \frac{\pi \hbar c}{2g}\left (\frac{1}{a} + \frac{1}{L-a}\right )(na - M - \varphi)^{2}
\end{eqnarray}
where we specified to the case of the Dirichlet wall at $x = L$ (which explains the third "D" in the "DDD" superscript) and employed the continuity of the field variable across the partition at $x = a$.  Since the latter is slightly transparent to particles, the integer parameter $M$ can 
adjust to minimize this expression for any chosen value of $a$.  The result is that the
ground-state energy of the system is an 
oscillatory function of the position of the partition. The usual Casimir effect is the lower 
envelope of the function, while particle discreteness contributes parabolic cusps.  
It should be noted that the scale of the discreteness effect is comparable to the
contribution from the zero-point motions.  The conclusion (\ref{EDDDgeneral}) is also applicable to the geometry of Fig.1b;  in the $a \ll L - a$ limit it reduces to the  Casimir interaction between two strong impurities placed in an infinite Luttinger liquid \cite{Zwerger}. 

\section{Microscopic analysis of spinless free fermions}

In the following subsections we will calculate the ground-state energy for a
model of non-interacting spinless fermions.  Since the long-wavelength excitations
of this system are described by a harmonic theory, the finite-size (Casmir) effects are captured by the continuum theory as described in previous Section.  Here we will be able to see how these macroscopic conclusions emerge from a completely different microscopic analyis of the exactly-solvable model.  Naturally this will not have the generality of phenomenological theory, however  all the details, including the phase of the Casimir oscillation $\varphi$ in Eq.(\ref{EDDDgeneral}) will be determined.

There is again an issue of boundary conditions imposed on the many-body wavefunction 
(hereafter indicated by the letters D or N) .  The (D) Dirichlet boundary condition 
for the fermion
wavefunction implies that no particles can reach the boundary, which is consistent with the Dirichlet condition imposed on the collective variable $u(x,t)$ (i.e. $u = 0$ at the boundary).
However, the Neumann condition for the collective variable ($\partial u / \partial x = 0$) allows
a current through the boundary, while the (N) Neumann condition for the wavefunction does not.  At the same time the (N) Neumann condition for the wavefunction is not inconsistent with the Dirichlet condition imposed on the collective variable.  
These statements representing plausible guesses will be verified below;  they imply that the conclusions derived from microscopic analysis of the Dirichlet and Neumann geometries can be only compared with the macroscopic Dirichlet results (\ref{ocasimir}) and (\ref{EDDDgeneral}).

\subsection{Segment and ring geometries}

 Before directly addressing the geometries of Fig.1 we find it useful to start with 
simpler segment and ring arrangements which may be viewed as reference states for those in 
Fig. 1. Here the Casimir interaction can be extracted as a finite-size correction to the 
bulk ground-state energy \cite{conformal}.

It is intuitively clear and will be made explicit below that the rearrangements of the density profile caused by either Dirichlet or Neumann boundary condition represent strong perturbations which cannot be correctly captured within the harmonic theory (\ref{action}).

\subsubsection{Like boundary conditions}

First consider $M$ fermions inside an impenetrable box of length $a$. This corresponds to 
the "D" boundary conditions enforced at the segment ends. The normalized single-particle 
wavefunctions for this problem are well-known:
\begin{equation}
\label{wavefunction}
\psi_{l}(x) = \left (\frac{2}{a}\right )^{1/2} \sin k_{l}x
\end{equation} 
where the allowed wave numbers satisfy the condition $k_{l} = \pi l/a$, $l = 1, 2, 3, ...$. 
The ground state is built by sequential occupation of the single-particle states beginning 
from $l = 1$ and ending with $l = M$. The resulting expressions for the ground-state energy 
\begin{eqnarray}
\label{eofma}
&E_{DD}(M, a)& = \sum_{l = 1}^{M}\frac{\hbar^{2}k_{l}^{2}}{2m} \nonumber\\
& = &\frac{\pi^{2}\hbar^{2}}{6m} \left (\frac{M + \frac{1}{2}}{a}\right )^{3}a - \frac{\pi^{2}\hbar^{2}}{24ma} \frac{M + \frac{1}{2}}{a}
\end{eqnarray} 
and the particle density inside the box
\begin{equation}
\label{density}
\rho_{DD}(x) = \sum_{l = 1}^{M} \psi_{l}^{2}(x) = \frac{M + \frac{1}{2}}{a} - \frac{\sin\left (2\pi(M + \frac{1}{2})x/a\right )}{2a\sin(\pi x/a)}
\end{equation}
are well-known \cite{Peierls}. Since the latter is depleted near the box walls, the average 
bulk density, as evidenced by Eq.(\ref{density}), is $(M + 1/2)/a$ rather than the naively 
expected $M/a$. Swiatecki \cite{Swiatecki} emphasized the importance of this fact in 
separating the bulk contribution to the ground-state energy from those due to the boundaries. 
We made this explicit in Eq.(\ref{eofma}) which has a form typical of the ground-state energy 
of a one-dimensional quantum system confined to a segment of length $a$ \cite{conformal}: 
The first term gives the bulk energy of the spinless free Fermi gas of density 
$n = (M + 1/2)/a$ while the second term, $-\pi^{2}\hbar^{2}n/(24ma) = - \pi \hbar c/(24a)$, 
represents the Casimir effect \cite{conformal}. 

If both ends of the segment enforce "N" boundary conditions, the single-particle ground-state has zero energy and is characterized by a constant wave function, $\psi_{0}(x) = 1/\sqrt{a}$. The excited states are of the form, $\psi_{l}(x) = (2/a)^{1/2}\cos k _{l}x$, and the allowed wave numbers are determined by $k_{l} = \pi l/a$, $l = 1, 2, 3, ...$. It is then straightforward to realize that the ground-state energy of the $M$-fermion segment follows from Eq.(\ref{eofma}) by substituting $M - 1$ for $M$: 
\begin{equation}
\label{freeEsegment }
E_{NN}(M, a)= \frac{\pi^{2}\hbar^{2}}{6m} \left (\frac{M - \frac{1}{2}}{a}\right )^{3}a - \frac{\pi^{2}\hbar^{2}}{24ma} \frac{M - \frac{1}{2}}{a}
\end{equation}
The particle density is then given by
\begin{equation}
\label{freedensity }
\rho_{NN}(x) = \sum_{l = 0}^{M-1} \psi_{l}^{2}(x) = \frac{M - \frac{1}{2}}{a} 
+ \frac {\sin(2\pi x(M - \frac{1}{2})/a)}{2a\sin(\pi x/a)}
\end{equation} 
Eq.(\ref{freedensity }) makes obvious that fact that "N" boundary conditions promote 
particle accumulation 
near the segment ends where the density reaches the maximal value of $(2M - 1)/a$. 
This is to be expected as the "N" boundary condition can be implemented via an 
attractive potential localized at the boundary. As a result the bulk density 
$n = (M - 1/2)/a$ is smaller than $M/a$. This is consistent with the expression for the 
ground-state energy (\ref{freeEsegment }) which implies an attractive Casimir 
interaction of the expected $- \pi \hbar c/(24a)$ form \cite{conformal}.

Although Eqs.(\ref{eofma}) and (\ref{freeEsegment }) do not explicitly address the 
geometry of Fig. 1a to which Eq.(\ref{ocasimir}) is supposed to be applicable, they 
certainly imply plausibility of Eq.(\ref{ocasimir}) for like boundary conditions. 

\subsubsection{DN  boundary conditions}

As a next step consider $M$ fermions belonging to a segment of length $a$ with the "D" boundary condition at $x = 0$ and "N" boundary condition at $x = a$. Then the single-particle wave functions are still given by Eq.(\ref{wavefunction}) while the wave numbers selected by the "N" boundary condition at $x = a$ satisfy $k_{l} = \pi (l - 1/2)/a$, $l = 1, 2, 3, ...$. As a result the ground-state energy and the particle density inside the segment are given by 
\begin{equation}
\label{unlikeEsegment }
E_{DN}(M, a) = \sum_{l = 1}^{M}\frac{\hbar^{2}k_{l}^{2}}{2m} = \frac{\pi^{2}\hbar^{2}}{6m}\left (\frac{M}{a} \right )^{3}a - \frac{\pi^{2} \hbar^{2}}{24ma}\frac{M}{a}
\end{equation} 
and
\begin{equation}
\label{ unlikedensity}
\rho_{DN}(x) = \sum_{l = 1}^{M} \psi_{l}^{2}(x) = \frac{M}{a} - \frac{\sin(2\pi Mx/a)}{2a\sin(\pi x/a)},
\end{equation}
respectively. The latter vanishes at the "D" end $x = 0$ and reaches the maximal value 
of $2M/a$ at the "N" end $x = a$. As evidenced by Eq.(\ref{ unlikedensity}), the 
particles pushed away from the "D" end accumulate at the "N" end, and as a result the 
bulk density $n$ takes on the naively expected $M/a$ value. Then 
Eq.(\ref{unlikeEsegment }) implies an attractive Casimir interaction, 
$- \pi \hbar c/(24a)$, of exactly the same form as was found for like boundary 
conditions (see Eqs.(\ref{eofma}) and (\ref{freeEsegment })).   

\subsubsection{Periodic boundary conditions}

Now assume that $M$ fermions are confined to a circle of circumference $a$. For odd M the 
single-particle wavefunctions for this problem are of the 
$\exp i k_{l}x$ form with the wavenumbers satisfying the condition 
$k_{l} = 2\pi l/a, l = 0, \pm 1, \pm 2, ...$. The ground state is built by placing one of the 
particles into the $l = 0$ state while the remaining $M - 1$ fermions symmetrically occupy 
the states beginning from $l = \pm 1$ and ending with $l = \pm (M - 1)/2$. 
The ground-state energy is thus given by
\begin{equation}
\label{eofmaring}
E_{periodic}(M, a) = \frac{\pi^{2}\hbar^{2}}{6m}\left (\frac{M}{a} \right )^{3}a - \frac{\pi^{2} \hbar^{2}}{6ma}\frac{M}{a} .
\end{equation}
Repeating the construction for even M leads to the same result. Eq. (\ref{eofmaring}) has a structure similar to Eqs.(\ref{eofma}), (\ref{freeEsegment }) and (\ref{unlikeEsegment }): the first term gives the bulk energy of spinless gas of free fermions of density $n = M/a$ (which is everywhere uniform) while the second term, $-\pi^{2}\hbar^{2}n/(6ma) = - \pi \hbar c/(6a)$, represents the Casimir effect. The magnitude of the former in the ring geometry is four times larger than that for the segment \cite{conformal}. 

\subsection{Geometries with three boundaries; continuum limit}

The geometries of Fig. 1 may be viewed as a result of insertion of one (a) or two (b) 
partitions inside a segment or a ring of length $L$ containing $N$ fermions. This brings 
up a qualitatively new aspect into the problem: the issue of particle discreteness. 
This was not relevant to the segment or ring geometries since the total particle number was 
fixed. Now as the partition (or partitions) is adiabatically displaced, particle exchange 
across it
has to be allowed -- otherwise instead of the Casimir effect one would be probing the bulk 
compressibility of the Fermi gas. The effects of particle discreteness are largest when the 
partition is nearly-impenetrable, i. e. it enforces the "D" boundary condition. 
In what follows we will only focus on this case. However for the geometry of Fig. 1a the 
boundary conditions at the segment ends could be either like (DD or NN) or unlike (DN or ND).

We begin with the case that the partition(s) and the walls enforce  Dirichlet  
constraints at all three boundaries. When the number of particles captured in each of the 
boxes in Fig.1 is large, these may be treated as continuous varibles. The equilibrium value
of a macroscopic variable (for example $M$) can be found by minimizing the ground-state 
energy, i. e. by requiring that 
$\partial ( E_{DD}(M,a) + E_{DD}(N - M, L - a))/\partial M = 0$. 
It is straightforward to verify that to leading order in $M, N - M \gg 1$ this gives
\begin{equation}
\label{minimum}
\frac{M + \frac{1}{2}}{a} = \frac{N - M + \frac{1}{2}}{L - a} \equiv \frac{N + 1}{L}
\end{equation} 
which, in view of Eq.(\ref{density}), can be recognized as the condition that the bulk densities in the two boxes be equal. Their common value of $(N + 1)/L$ is higher than $(N + 1/2)/L$ of the reference $N$-fermion segment of length $L$ or $N/L$ for the $N$-fermion ring of circumference $L$ because the partition(s) push the particles into the bulk. 

Combining Eqs. (\ref{eofma}) and (\ref{minimum} ) we find that the total energy for the geometries of Fig. 1, $E_{DDD} = E_{DD}(M,a) + E_{DD}(N - M, L - a)$, is given by
\begin{equation}
\label{contcasimira}
E_{DDD} = \frac{\pi^{2}\hbar^{2}n^{3}}{6m}L - \frac{\pi \hbar c}{24} \left (\frac{1}{a} + \frac{1}{L - a}\right )
\end{equation}
where $n = (N + 1)/L$ and we used the fact that $\pi \hbar (N + 1)/(mL) = \pi\hbar n/m = c$ is the sound velocity. 

We observe that the first term of Eq.(\ref{contcasimira}) is of order $O(L)$ and 
represents the bulk energy of a Fermi gas of density $n = (N + 1)/L$. The remaining $a$-dependent  terms combined 
which are of order $O(a^{-1})$ and $O((L-a)^{-1})$ are identical to the result based on regularization,
 thus supporting Eq.(\ref{ocasimir}). 
Since the $O(1)$ term is absent, we conclude that there is zero self-energy
associated with a single partition. This is consistent with Eq.(\ref{eofma})
which also predicts zero self-energy but in variance 
with the earlier conclusions of Swiatecki \cite {Swiatecki} and Peierls \cite {Peierls}.

The analysis of the NDN and DDN geometries (Fig. 1a) leads to the same result (\ref{contcasimira}) with $n = N/L$ and $n = (N + \frac{1}{2})/L$, respectively.

\subsection{Geometries with three boundaries; effects of particle discreteness}

The main conclusion of the previous analysis is that when the 
particle discreteness is ignored, the Casimir interaction is universally given by Eq.(\ref{ocasimir}). Specifically, for the segment geometry of Fig. 1a it is insensitive to the boundary conditions enforced at the segment ends (provided they form any combination of "N" and/or "D" constraints).  As a next step we improve the analysis by taking into consideration the particle discreteness. 

\subsubsection{DDD boundary conditions: nearly-impenetrable partition(s) and "D" walls, Fig. 1}

In order to capture only the $a$-dependent contributions the Casimir energy for this case will be defined as
\begin{eqnarray}
\label{meso}
E_{C, DDD} &=& E_{DD}(M, a) + E_{DD}(N - M, L - a)\nonumber\\ &-& \frac{\pi^{2}\hbar^{2}(N + 1)^{3}}{6mL^{2}} 
\end{eqnarray} 
and it is assumed that for given $a$ an {\em integer} $M$ is chosen to minimize (\ref{meso}). We already demonstrated (see Eq.(\ref{contcasimira})) that if the discrete character of $M$ is ignored (and $M, N - M \gg 1$) then $E_{C, DDD}$ reduces to Eq.(\ref{ocasimir}). Since the latter is the result of unconstrained minimization with respect to $M$, while $M$ must be an integer in the fermion model, we conclude that the effect cannot be smaller than its continuum approximation. In other words, the continuum approximation is the lower envelope of the interaction given by Eq. (\ref{meso}).

\textbf{Microscopic consideration.}  In order to understand the $E_{C, DDD}(a)$ dependence (shown in Fig.2) , let us begin with a very small size $a$ for one of the boxes in Fig. 1. Then the energy cost of localizing even one particle inside the box is very large; as a result this box is empty and all the particles reside inside the second box of size $L - a$. This continues to be the case as $a$ adiabatically increases as long as the latter remains sufficiently small. Thus the $E_{C, DDD}(a)$ dependence is given by the $M = 0$ branch of Eq.(\ref{meso}) - it is an increasing function of $a$ reflecting the compression of $N$ fermions captured inside the box of decreasing size $L - a$. As $a \rightarrow 0$ the force exerted on the partition (Fig.1a) or the interaction force between two partitions (Fig. 1b), $F = -\partial E_{C, DDD}/\partial a$, approaches $-\pi^{2}\hbar^{2}(2N^{3} + 3N^{2} + N)/(6mL^{3})$ which remains finite in the limit that $N, L \rightarrow \infty$ with $N/L$ fixed. The $M = 0$ regime obtains until the empty box becomes so large that the energy cost of localizing a fermion inside it equals the strain energy released when a particle tunnels out of the compressed box. At that instant, $E_{DD}(N, L - a) = E_{DD}(1, a) + E_{DD}(N - 1, L - a)$, the $E_{C, DDD}(a)$ dependence switches to the $M = 1$ branch of Eq.(\ref{meso}). 

As $a$ continues to increase, the function $E_{C, DDD}(a)$ first decreases because the associated falloff of the energy of localization of a fermion inside the box of size $a$ prevails over the increase of energy of the compressed $(N - 1)$-fermion box. This trend however is eventually reversed and the energy cost of compression begins to dominate over the energy gain of delocalization. As a result, the $E_{C, DDD}(a)$ dependence has a minimum. 

Upon further increase of $a$, a point is reached when the single-fermion box is too large, while its $(N - 1)$-fermion counterpart is too compressed, so that it is energetically favorable for a particle to tunnel into the box of size $a$. At that moment, $E_{DD}(1, a) + E_{DD}(N - 1, L - a) = E_{DD}(2, a) + E_{DD}(N - 2, L - a)$, the $E_{C, DDD}(a)$ dependence switches to the $M = 2$ branch of Eq.(\ref{meso}).   The competition between delocalization and compression periodically resolved by tunneling keeps recurring as $a$ further increases, and the function $E_{C, DDD}(a)$ switches through the $M = 3, 4, ..., N$ branches of Eq.(\ref{meso}). At the transition points between the neighboring branches the ground state is degenerate, $E_{DD}(M, a) + E_{DD}(N - M, L - a) = E_{DD}(M + 1, a) + E_{DD}(N - M - 1, L - a)$. All but $M = 0$ and $M = N$ branches of $E_{C, DDD}(a)$ have minima where the Casimir force vanishes; their loci should be well-approximated by the continuum result (\ref{ocasimir}). 

To summarize, the Casimir interaction (\ref{meso}) is an oscillatory piecewise-continuous function of $a$ whose maxima, the points of force discontinuity, correspond to resonant tunneling between the boxes in Fig. 1. This agrees with previous conclusions \cite{Volovik2,Zwerger,lattice, Anfuso}. 

\textbf{Macroscopic limit.}  When there are large numbers of particles on either side of the partition, the function
$E_{C, DDD}(a)$ can be accurately characterized by mathematically simpler
 expressions. 
The minima of the Casimir energy Eq. (\ref{meso}) occur when the densities on
the two sides of the partition are equal.
This is the condition already given in Eq.(\ref{minimum}), which can be restated to imply that
the loci of the zeros of the Casimir force, measured in units of the bulk interparticle spacing $n^{-1} = L /(N + 1)$,
are half-integers:
\begin{equation}
\label{am}
\frac{N + 1}{L}a_{M} \equiv na_{M}= M + \frac{1}{2}
\end{equation}
where $M$ is the integer that labels the $M$-th minimum of the Casimir interaction.

The transition points, $a_{M \rightarrow M +1}$, between the $M$-th and $(M + 1)$-th branches of the function (\ref{meso}) follow from the condition of degeneracy, $E_{DD}(M, a) + E_{DD}(N - M, L - a) = E_{DD}(M + 1, a) + E_{DD}(N - M - 1, L - a)$, Taylor expanded up to second order in $\Delta M = 1$. We find that in units of the bulk interparticle spacing, the transition points are given by
\begin{equation}
\label{amm+1}
\frac{N + 1}{L}a_{M \rightarrow M + 1} \equiv na_{M\rightarrow M + 1} = M + 1,
\end{equation}
i. e. they are at integer values. Not surprisingly, the distance between the nearest minima of the Casimir interaction (\ref{am}) is the same as the distance between nearest maxima (\ref{amm+1}) and equal the bulk interparticle spacing. At the same time the distance between a transition point (\ref{amm+1}) and two nearest minima of (\ref{meso}) is half interparticle spacing. 

Since the loci of the minima of the Casimir interaction (\ref{am}) are symmetrically sandwiched between the transition points (\ref{amm+1}), the $M$-th branch of the function (\ref{meso}) may be approximated by its second order Taylor expansion around $a = a_{M}$:
\begin{eqnarray}
\label{parabola}
E_{C, DDD}(a) &\approx& \frac{\pi^{2}\hbar^{2}n^{2}}{24m}\left (\frac{1}{M + \frac{1}{2}} + \frac{1}{N - M + \frac{1}{2}}\right )\nonumber\\&\times& \left (-1 + 12(na - M - \frac{1}{2})^{2}\right )\nonumber\\ &\approx&- \frac{\pi\hbar c}{24}\left (\frac{1}{a} + \frac{1}{L - a}\right )\nonumber \\
&\times& \left (1 - 12(na - M - \frac{1}{2})^{2}\right )
\end{eqnarray}
which is valid when $na$ is positioned between nearest integers
\begin{equation}
\label{range}
M \leqslant na < M + 1
\end{equation}

In the second representation of Eq.(\ref{parabola}) the difference between $a_{M}$ and $a$ was neglected in the coefficients of the Taylor expansion which is a valid approximation within the range (\ref{range}). This allows us to see that Eq.(\ref{parabola}) is identical to the macroscopic result (\ref{EDDDgeneral}) with $g=1$ (non-interacting fermions) and  $\varphi = 1/2$.
 
\begin{figure}
\includegraphics[width=1.0\columnwidth,keepaspectratio]{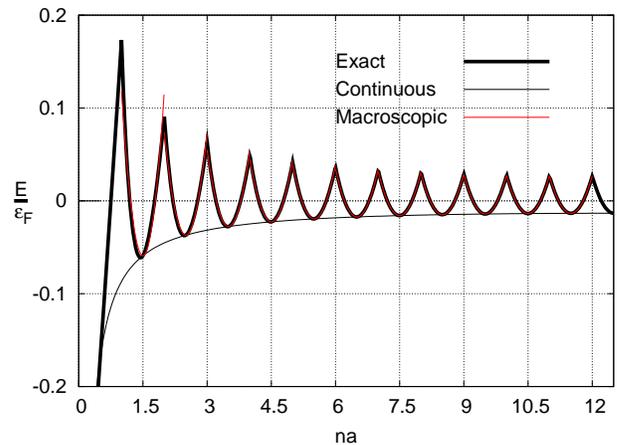} 
\caption{(Color online) Casimir interaction for "D" boundary conditions enforced at all boundaries, Eq.(\ref{meso}), and its continuum, Eq.(\ref{ocasimir}) and macroscopic, Eq.(\ref{parabola}), approximations in units of the bulk Fermi energy $\epsilon_{F} = \pi^{2}\hbar^{2}n^{2}/2m$, as a function of dimensionless position of the partition $na$ (Fig.1a) or distance between the partitions (Fig.1b) in the $0 \leqslant na \leqslant (N + 1)/2$ range for $N = 24$ 
fermions.  The macroscopic theory is indistinguishable from the exact result, except at
very small distances.}
\end{figure}

As implied by Eq.(\ref{parabola}), the oscillations of the Casimir force $F = -\partial E_{C, DDD}/\partial a$ are confined between upper and lower envelopes whose macroscopic limit is given by
\begin{equation}
\label{ envelope}
F_{envelope} = \pm \frac{\pi^{2}\hbar^{2}n^{2}}{2m}\left (\frac{1}{a} + \frac{1}{L-a}\right )
\end{equation}
The amplitude of this oscillation, $2|F_{envelope}|$, is the magnitude of the jump of the Casimir force at the point of resonant tunneling across the partition(s). This result was first given by Volovik \cite{Volovik2}.
We should compare this with the magnitude of the force computed from the
continuum theory: according to Eq. (\ref{ocasimir}), it varies as $1/a^{2}$.
Then in the $1 \ll M \ll N$ limit,  the force exerted on the
partition (Fig.1a) or the interaction force between two partitions (Fig.1b) is
a factor of $M$ larger than its continuum counterpart \cite{Volovik2}.

For the purposes of comparing of the continuum, Eq.(\ref{ocasimir}), and macroscopic, Eq.(\ref{parabola}), approximations to the exact Casimir interaction, Eq.(\ref{meso}), it is natural to measure the energy in units of the bulk Fermi energy $\epsilon_{F} = \pi^{2}\hbar^{2}n^{2}/2m$, while the position of the partition $a$ (Fig.1a) or the distance between the partitions (Fig.1b) is measured in units of the bulk interparticle spacing $n^{-1}$. Then the variable $na$ varies between zero and $N + 1$; in view of the reflection symmetry about $a = L/2$ we only need to look at the $0 \leqslant na \leqslant (N + 1)/2$ range. The result is plotted in Fig. 2 for the case $N = 24$.

We note that even for the small-$M$ branches of the unapproximated interaction (\ref{meso}) the lower envelope is well-appproximated by the continuum result (\ref{ocasimir}). The macroscopic result (Eq.(\ref{parabola})) represents an excellent approximation to the interaction (\ref{meso}). Although the $M = 0$ branch is not reproduced, the $M = 1$ branch is approximated fairly well, and Eqs.(\ref{meso}) and Eq.(\ref{parabola}) are hardly distinguishable for $M \geqslant 3$. We also note that for an even number of particles $N$ there is a minimum of the Casimir interaction in the middle of the system at $na = (N + 1)/2$, as demonstrated in Fig.2. Had we chosen $N$ odd, there would be instead a maximum at $na = (N + 1)/2$.

To judge the accuracy of the various approximations it is more relevant to look at the experimentally measurable Casimir force, $F = -\partial E_{C, DDD}/\partial a$, rather than the interaction. The former (measured in units of $n\epsilon_{F}$) is plotted as a function of $na$ in Fig. 3 with the same choice of $N = 24$ fermions as in Fig. 2.
\begin{figure}
\includegraphics[width=1.0\columnwidth,keepaspectratio]{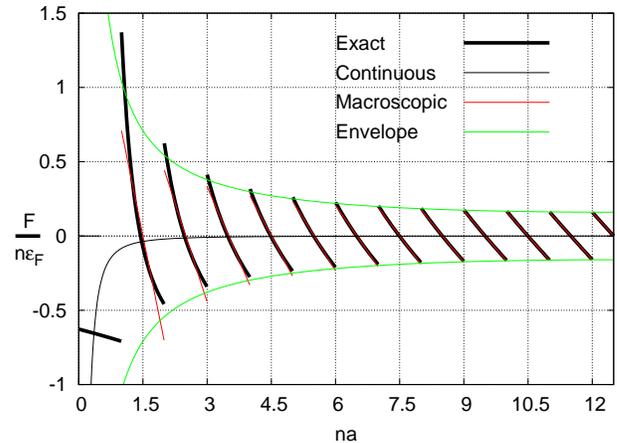} 
\caption{(Color online) Casimir force $F = -\partial E_{C, DDD}/\partial a$ and its approximations corresponding to the energy curves of Fig.2. The force is measured in units of $n\epsilon_{F}$. Additionally, the macroscopic limit of the upper and lower envelopes of the force, Eq.(\ref{ envelope}), is shown.
The "macroscopic" and "exact" curves almost entirely coincide, and give a force of oscillating
magnitude and sign that is generally large compared to the continuum result.}
\end{figure}
We now see that the continuum limit gives a very poor representation of the Casimir force. The macroscopic approximation, on the other hand, continues to be very accurate; for $M \geqslant 6$ it is hardly possible to distinguish it from the unapproximated result. The large $M$ (large $na$) part of the 
force plot resembles Figure 29.3 of Volovik \cite{Volovik2}. 

As anticipated in Section IIB, the deviation of the exact fermionic Casimir interaction from its continuum counterpart (\ref{ocasimir}) which is due the combined effect of particle discreteness and near impenetrability of the partition(s) can be understood in terms of the effective boundary condition at the partition(s) (\ref{boundarycondition}) imposed on the displacement field $u(x,t)$ of the continuum theory (\ref{action}).  For the DDD geometry this condition becomes
\begin{equation}
\label{DDDboundarycondition}
u(a,t) = a - \frac{M + \frac{1}{2}}{n}
\end{equation}
where we used $\varphi = 1/2$.  At the values of $a$ for which unconstrained minimization of the ground-state energy with respect to $M$ happens to give an integer $M$, the boundary condition in question is  "impenetrable". These $a$'s correspond to the minima of the Casimir interaction where the continuum approximation is accurate. Indeed combining Eqs.(\ref{am}) and (\ref{DDDboundarycondition}) we arrive at the Dirichlet condition $u(a,t) = 0$.  At any other value of $a$ the effective boundary condition interpolates between the "impenetrable" and "transparent" limits. The latter corresponds to half-integer $nu(a,t)$ as can be seen by combining Eqs. (\ref{amm+1}) and (\ref{DDDboundarycondition}).   The "transparency" is perfect at the points of resonant tunneling which are the maxima of the Casimir interaction; in the macroscopic limit their loci are implicit in Eq.(\ref{parabola}): 
\begin{equation}
\label{energyenvelope}
E_{envelope} = \frac{\pi \hbar c}{12} \left (\frac{1}{a} + \frac{1}{L - a}\right )
\end{equation}
The latter has the same sign as Eq.(\ref{unlike}) describing the case of the Neumann partition ($\partial u/\partial x = 0$) and the Dirichlet walls ($u = 0$) but a different magnitude. It is curious that the average of the lower (\ref{ocasimir}) and upper (\ref{energyenvelope}) envelopes of the interaction is given precisely by Eq.(\ref{unlike}).

The analysis of the Casimir interaction for the cases of "N" or mixed boundary conditions enforced at the segment ends is very similar to the one just presented; only a summary of the main results emphasizing the differences between various cases is provided below.   

\subsubsection{NDN boundary conditions: nearly-impenetrable partition, "N" walls, Fig. 1a}

The $a$-dependent part of the Casimir interaction is given by
\begin{equation}
\label{mesounlike }
E_{C, NDN}^{(a)} = E_{ND}(M, a) + E_{DN}(N - M, L - a)
- \frac{\pi^{2}\hbar^{2}N^{3}}{6mL^{2}} 
\end{equation}
In the macroscopic limit the loci of the zeros of the Casimir force measured in units of the bulk interparticle spacing $n^{-1} = L/N$ are integer:
\begin{equation}
\label{ amunlike}
\frac{N }{L}a_{M} \equiv na_{M}= M
\end{equation} 
In the same dimensionless units the transition points between the $M$-th and $(M+1)$-th branches of the function (\ref{mesounlike }) or, equivalently, the maxima of the Casimir interaction are given by half-integers:
\begin{equation}
\label{ amm+1unlike}
\frac{N}{L}a_{M\rightarrow M +1} \equiv na_{M\rightarrow M + 1}= M + \frac{1}{2}
\end{equation}
The $M$-th branch of the function (\ref{mesounlike }) may be approximated by its second order Taylor expansion around $a = a_{M}$:
\begin{eqnarray}
\label{parabolaunlike}
E_{C, NDN}^{(a)}(a) &\approx& \frac{\pi^{2}\hbar^{2}n^{2}}{24m}\left (\frac{1}{M} + \frac{1}{N - M}\right )\nonumber\\&\times& \left (-1 + 12(na - M)^{2}\right )\nonumber\\ &\approx&- \frac{\pi\hbar c}{24}\left (\frac{1}{a} + \frac{1}{L - a}\right )\nonumber \\
&\times& \left (1 - 12(na - M)^{2}\right ) 
\end{eqnarray}
which is valid when $na$ is positioned between nearest half-integers
\begin{equation}
\label{rangeunlike}
M - \frac{1}{2} \leqslant na < M + \frac{1}{2}
\end{equation}
Eqs.(\ref{mesounlike }) - (\ref{rangeunlike}) are counterparts of the set of Eqs.(\ref{meso}) - (\ref{range}) describing the macroscopic limit of the Casimir effect with all the boundaries subject to the "D" condition. In the macroscopic limit, the only difference between the two is the location of the extrema of the Casimir interaction: now the minima lying at integer values of $na$ are sandwiched between half-integer maxima of the Casimir energy. This is a manifestation of the long-range effect exerted by the ends of the segment.  We also note that Eq.(\ref{parabolaunlike}) coincides with the $g  = 1$, $\varphi = 0$ case of the phenomenological result (\ref{EDDDgeneral}). 
\begin{figure}
\includegraphics[width=1.0\columnwidth,keepaspectratio]{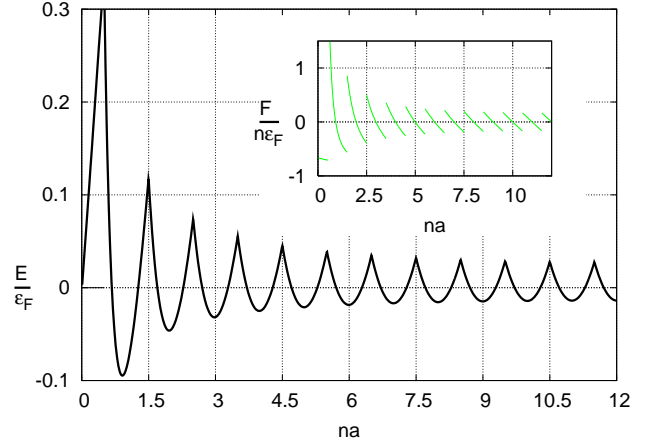} 
\caption{(Color online) Casimir interaction for nearly-impenetrable partition and "N" boundary conditions enforced at the segment ends, Eq.(\ref{mesounlike }), in units of bulk Fermi energy $\epsilon_{F}$ as a function of dimensionless position of the partition $na$ (Fig. 1a) in the $0 \leqslant na \leqslant N/2$ range for $N = 24$ fermions. The inset shows the plot of the corresponding Casimir force.}
\end{figure}

These conclusions are illustrated in Fig. 4 where the Casimir interaction, Eq.(\ref{mesounlike }), is plotted in dimensionless units of the Fermi energy $\epsilon_{F}$ as a function of the dimensionless position of the partition $na$ (Fig. 1a). Since now the bulk density is given by $n = N/L$, the variable $na$ varies between zero and $N$; in view of the reflection symmetry about $a = L/2$, only the $0 \leqslant na \leqslant N/2$ range is shown. The inset shows the Casimir force. To facilitate comparison with the previously studied case of "D" boundary conditions enforced at all boundaries (Figs. 2 and 3), we again show the case $N = 24$. It now becomes apparent that the bulk behavior is identical in the two cases, apart from a shift by a half along the $na$ axis. Additional differences overlooked in the macroscopic approximation are limited to the vicinity of small $na$; they represent manifestations of the distinction between the "D" and "N" wall boundary conditions enforced at $na =0$. 

\subsubsection{DDN boundary conditions: nearly-impenetrable partition, unlike boundary conditions at the walls, Fig. 1a}

The $a$-dependent part of the Casimir interaction is given by
\begin{eqnarray}
\label{mesoasymmetric}
E_{C, DDN}^{(a)}& =& E_{DD}(M,a) + E_{DN}(N - M, L - a)\nonumber\\
&-&\frac{\pi^{2}\hbar^{2}(N + \frac{1}{2})^{3}}{6mL^{2}} 
\end{eqnarray} 
This case is extremely similar to that of all three boundaries constrained to the "D" boundary condition (Section IIIC1). In the macroscopic limit it is impossible to tell the two apart even when the effects of particle discreteness are accounted for. The one slight difference (apart from the value of the bulk density $n= (N + 1/2)/L$) is that the $N - M + 1/2$ combination on the first line of Eq.(\ref{parabola}) has to be replaced with $N - M$. This difference however disappears in the second representation of Eq.(\ref{parabola}).  The macroscopic limit again coincides with the $g = 1$, $\varphi=1/2$ case of the phenomenological result (\ref{EDDDgeneral}). 
\begin{figure}
\includegraphics[width=1.0\columnwidth,keepaspectratio]{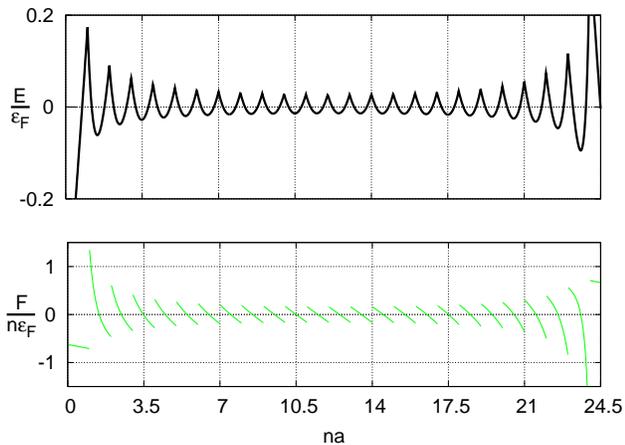} 
\caption{(Color online) Top: Casimir interaction for nearly-impenetrable partition with "D" and "N" boundary conditions enforced at the left and right segment ends, respectively, Eq.(\ref{mesoasymmetric}), in units of bulk Fermi energy $\epsilon_{F}$ as a function of dimensionless position of the partition $na$ (Fig. 1a) in the $0 \leqslant na \leqslant N + 1/2$ range for $N = 24$ fermions. Bottom: corresponding dimensionless Casimir force.}
\end{figure}

These observations are illustrated in Fig. 5 where the Casimir interaction (top), Eq.(\ref{mesoasymmetric}), is plotted in dimensionless units of the Fermi energy $\epsilon_{F}$ as a function of the dimensionless position of the partition $na$ (Fig. 1a). Since the bulk density is $(N + 1/2)/L$, the variable $na$ varies between zero and $N + 1/2$. Since reflection symmetry is lacking, the whole $0 \leqslant na \leqslant N + 1/2$ range is shown with the same choice of $N = 24$ fermions. The bottom plot shows the corresponding dimensionless Casimir force as a function of $na$. We notice that except for near a "N" wall $na = N + 1/2$, the energy and force plots are nearly identical to their counterparts, Figs. 2 and 3, with all boundaries subject to the "D" wall boundary conditions. On the other hand, the immediate vicinity of the right boundary, $na = N + 1/2$, is similar to the behavior near the "N" end, $na = 0$, in Fig. 4. 

For the cases of the mixed boundary conditions the macroscopic limit of the upper envelope of the Casimir interaction continues to be given by Eq.(\ref{energyenvelope}). Since this is the locus of the points of resonant tunneling, then we can  argue that for a "perfectly transparent" partition and any combination of "D" or "N" wall boundary conditions enforced at the segment ends, the Casimir interaction is given by Eq.(\ref{energyenvelope}). 

\section{Discussion}

In this paper we employed a model of spinless free fermions in one dimension to complement the macroscopic approach to the Casimir effect where it is derived from a regularized low-energy harmonic field theory. In the segment geometry with like (either DD or NN) or periodic boundary conditions when the discreteness of the underlying particles is irrelevant, we found an attractive Casimir interaction of the $-\pi\hbar c/(24a)$ or $-\pi\hbar c/(6a)$ form, respectively, which is in agreement with well-known results of the conformal field theory \cite{conformal}. Additionally we found an attractive Casimir interaction of the $-\pi\hbar c/(24a)$ form for the case of unlike (DN or ND) boundary conditions. 

When a nearly-impenetrable partition is inserted inside the segment, a redistribution of the particles takes place so as to minimize the system energy; the discreteness of the fermions plays a major role in determining the Casimir force exerted on the partition. However if the particle discreteness is ignored, then for any combination of the D and N boundary conditions enforced at the segment ends, the lower envelope of the Casimir interaction is found to be universally given by Eq.(\ref{ocasimir}).

Particle discreteness gives rise to new effects beyond the conventional Casimir force,
as first noted by Volovik  \cite{Volovik2}.  However, we disagree with
his conclusion that for the example he proposed (and we worked out) 
"the  Casimir vacuum pressure is not within the responsibility of the effective theory, and the microscopic (trans-Planckian) physics must be evoked" \cite{Volovik4}.  This is demonstrated by showing how the effects of the particle discreteness can be understood macroscopically;  the result, Eq.(\ref{EDDDgeneral}), is in agreement with microscopic analysis of spinless free fermions.

The arguments that led to Eq.(\ref{EDDDgeneral}) highlight the aspects of the Casimir effect 
in one-dimensional quantum liquids which are obscured in the exact microscopical treatment 
of non-interacting fermions:

First, the effect of particle discreteness accumulated in the $(na - M - \varphi)^{2}$  term 
of Eqs.(\ref{lagrangian}), (\ref{portion}) and (\ref{EDDDgeneral}) is of nearly-classical 
origin because the magnitude of the effect does not explicitly depend on Planck's constant 
$\hbar$.  This fact is not obvious from the exact analysis of free fermions because the 
latter system does not have a classical limit; the sound velocity $c$ in a free-fermion gas 
is determined by Planck's constant $\hbar$.  The only quantity which can be due to quantum 
effects is the phase of the Casimir oscillation set by the parameter $\varphi$.

Second, the oscillatory part of the Casimir effect is not an artifact of the fermionic 
model, as the low-energy theory (\ref{action}) is applicable both to interacting fermions and 
repulsive bosons.

Third, the dependence of the effect on the square of $na - M - \varphi$ in Eqs.(\ref{parabola}) and (\ref{parabolaunlike}) 
directly demonstrates that we are dealing with a phenomenon which is within the reach of the 
harmonic theory.  Moreover all the parameters entering Eq.(\ref{EDDDgeneral}), even the 
particle density $n$ (which here arises from the reciprocal of the interparticle spacing 
rather than the wavevector cutoff), are macroscopic.  Thus there is no need to invoke 
microscopic physics to understand the  effect.  The only exception to this might involve the 
parameter $\varphi$ which our macroscopic analysis did not determine.  However there is an 
argument that concludes that for arbitrary harmonic liquid in the DDD geometry one has
$\varphi = 1/2$ in the $L \rightarrow \infty$ limit \cite{Zwerger}.  The same value was 
found rigorously in this work for non-interacting fermions and an arbitrary relationship 
between $a$ and $L$.  Although more work is needed, it seems plausible that for an arbitrary 
harmonic liquid occupying the Dirichlet segment of length $L$ with a nearly-impenetrable 
partition at $x = a$ one has $\varphi = 1/2$.

We see that the continuum contributions to the Casimir effect agree with corresponding regularization-based conclusions, Eq.(\ref{ocasimir});  the oscillatory part of the effect, up to its phase, is common to all the geometries studied, and is correctly described by the macroscopic result (\ref{EDDDgeneral}).     

To summarize,  we have demonstrated that in the macroscopic limit, the results of the rigorous analysis of the Casimir effect in a one-dimensional gas of spinless free fermions are in agreement with the predictions of the bosonic low-energy field theory (\ref{action}) if the walls are of the Dirichlet type and if the particle discreteness is properly accounted for.  The latter implements the physical fact of near impenetrability which  means the number of particles captured on either side of the partition is quantized.  The difference between Eqs.(\ref{ocasimir}) and (\ref{EDDDgeneral}) is due to the particle discreteness effect;  the comparison of Eqs.(\ref{EDDDgeneral}) and (\ref{parabola}) (describing specific model under the same physical conditions) shows that the Casimir effect for a fermionic system is properly described by a macroscopic theory.  

Although it seems unlikely that the Neumann case will be any different, it would still be interesting to establish what kind of the wavefunction boundary condition corresponds to the Neumann boundary condition imposed on the collective variables, and to repeat our analysis for this case.  This is important because (ignoring the effects of the particle discreteness) all the examples we studied exhibited only attractive Casimir interaction.  Additionally, it would be very useful to be able to derive (rather than guess) macroscopic boundary conditions starting from microscopic theory.  

\section{Acknowledgements}

We thank F. Anfuso and A. Bulgac for bringing  their work on related topics to our attention.  

This work was supported by the Thomas F. Jeffress and Kate Miller Jeffress Memorial Trust.

\end{document}